\documentclass[12pt,a4paper,final]{iopart}

\usepackage{iopams}

\usepackage{graphicx}
\usepackage{dcolumn}
\usepackage[usenames,dvipsnames,svgnames,table]{xcolor}
\usepackage[colorlinks=true,
            linkcolor=blue,
            urlcolor=black,
            citecolor=blue]{hyperref}
\usepackage{bm}
\usepackage{enumerate}
\usepackage{lipsum}
\usepackage{epstopdf}
\usepackage{float}
\usepackage[caption = false]{subfig}
\usepackage{tabularx}
\usepackage{color}

\begin{document}

\title{ Electron-correlation study of Y III-Tc VII ions using a relativistic coupled-cluster theory }

\author[cor1]{Arghya Das$^{1}$, Anal Bhowmik$^{1}$,  Narendra Nath  Dutta$^2$, Sonjoy Majumder$^1$}
\address{$^1$Department of Physics, Indian Institute of Technology Kharagpur, Kharagpur-721302, India.}

\address{$^2$Department of Chemistry and Biochemistry, Auburn University,  Alabama-36849, USA.}
\eads{\mailto{arghyadas.1988@gmail.com}}
\eads{\mailto{analbhowmik@phy.iitkgp.ernet.in}}
\eads{\mailto{sonjoym@phy.iitkgp.ernet.in}}

\begin{abstract}
Spectroscopic properties, useful for plasma diagnostics and astrophysics, of a few rubidium-like ions are studied here. We choose one of the simplest, but correlationally  challenging series where $d-$ and $f-$ orbitals are present in the core and/or valence shells with $4d$ $^2D_{3/2}$ as the ground state. We study different correlation characteristics of this series and make precise calculations of electronic structure and rates of electromagnetic transitions. Our calculated  lifetimes and transition rates are compared with other available experimental and theoretical values. Radiative rates of vacuum ultra-violet electromagnetic transitions of the long lived Tc$^{6+}$ ion, useful in several areas of physics and  chemistry, are estimated. To the best of our knowledge, there is no literature  for most of these transitions. 

\end{abstract}

\maketitle

\section{INTRODUCTION}

Rubidium-like ions are favorable candidates as testing ground for high precision spectroscopic measurements and accurate theoretical  calculations due to their relatively simple, but highly correlated,  electronic structure.  In the present theoretical work, we have considered a few ions of the Rb isoelectronic sequence (Y III, Zr IV, Nb V, Mo VI and Tc VII) having the ground state configuration $[{Kr}]4p^64d^1$.  Recent investigation has reported the evidence of the presence of Y III and Zr IV in the spectrum of a subdwarf B star \cite{Naslim2011}.  Being a metastable state, the first excited state $4^2D_\frac{5}{2}$ has potential applications in astrophysics and plasma diagnostics \cite{Feldman1981, Fournier1997}. The forbidden transitions between the $4^2D_\frac{3}{2}$ and  $4^2D_\frac{5}{2}$ states of these ions can have application in quantum memory devices due to substantial lifetimes of the $4^2D_\frac{5}{2}$ state \cite{Heshami2016}. The fine structure splittings of $4^2D$ state for the first few members of this sequence are quite large (724 cm$^{-1}$ for Y III and 1250 cm$^{-1}$ for Zr IV). Another interesting member of this sequence is Tc$^{6+}$ ion. Merrill detected the presence of $^{97}$Tc in a red giant star in 1952 \cite{Merrill1952}.  All the three long lived isotopes ($^{97,98,99}$Tc) of tecnetium  \cite{Palmeri2005} are abundant in various scenarios of star evolution. A star can produce this element through nuclear fusion or neucleosynthesis. And its abundance reflects the evolution process of the star \cite{van1999}. Recently, Werner \textit{et al.} \cite{Werner2015} analyzed abundances of technetium ions in ultraviolet spectra of hot white dwarfs and showed that Tc$^{6+}$ may be present there. The Tc$^+$ ion, relevant for astronomical observations, had been studied by Palmeri \textit{et al.} \cite{Palmeri2007}. However, very few electronic structure data for Tc VII are found in literature. The observations of neutral\cite{Uttenthaler2007} and ionized technetium\cite{Werner2015} suggest the significance of our calculations on oscillator strength of the Tc$^{6+}$ ion.
  
         Lifetimes and oscillator strengths for a few allowed transitions of rubidium-like ions were calculated by Zilitis \cite{Zilitis2007, Zilitis2009} using the Dirac-Fock method. Most recently, core-polarization augmented Dirac-Fock oscillator strengths have been studied for this ion sequence in the vicinity of the \textit{d} orbital collapse region  \cite{Migdalek2016}. Safronova and Safronova \cite{Safronova2013} calculated transition properties of various allowed and forbidden transitions  for Y III using a relativistic all-order many-body perturbation method (RMBPT). Measurements on the lifetimes of a few low-lying states of Y III were conducted using the time-revolved laser-induced fluorescence method by Bi$\acute{e}$mont \textit{et al.} \cite{Biemont2011}. In adition, a number of theoretical calculations and experimental measurements were performed in the last two decades on various atomic properties of different Rb-like ions, which are discussed in detail in Ref.\cite{Zilitis2007,Maniak1994}.

 The aim of this paper is three fold:  a) to calculate the ionization energies (IEs) of few low-lying states and estimate the electric dipole oscillator strengths among them for YIII-Tc VII ions, b) to study the nature of different correlation for this sequence, c) to make correlation exhaustive estimations of allowed and forbidden transition amplitudes for a few low-lying states and their contributions to accurate estimation of lifetimes of the states. These correlation studies allow us to estimate the accuracy of our calculations for Tc VII. We have  employed the relativistic coupled-cluster (RCC) theory \cite{Dixit2008, Dutta2014, Dutta2016,  Bishop1991, Anal2017} with cluster operators corresponding to single, double and partially triple excitations for the estimations of the correlations. This is one of the most accurate many-body methods to calculate the correlated atomic properties.

\section{THEORY}

Here our primary aim is to solve the energy eigenvalue equation, i.e., 
$H|\psi_v\rangle=E_v|\psi_v\rangle$ corresponding to the Dirac-Coulomb Hamiltonian \cite{Dutta2016, Bishop1987, Lindgren1987,   Dixit2007}
\begin{equation}
 H=\sum_{i=1}^{N}\left(c{\bf \alpha}_{i}\cdot{\bf p}_{i}+(\beta_{i}-1) c^{2}+V_{\mathrm{nuc}}({\bf r}_{i})+\sum_{j<i}\left( \frac{1}{{\bf r}_{ij}}\right)\right).
\end{equation}
Here, $|\psi_v\rangle$ is a correlated wave function of a single valence ion, $v$ represents the orbital containing the valence electron. Using the coupled-cluster theory \cite{Lindgren1987}, one can express this correlated wave function as 
  \begin{equation}
  |\psi_v\rangle=e^T\{1+S_v\}|\phi_v\rangle.
  \end{equation}
 The state $|\phi_v\rangle$ has been generated at the Dirac-Fock level with $V^{N-1}$ potential ($N$ is the total number of electrons) approximation using Koopman's theorem \cite{Szabo1996}. Here $T $ and $S_v$ are the closed-shell and open-shell cluster operators, respectively. In the present work, the cluster operator $T$ considers single and double excitations to a finite set of virtual orbitals from the core orbitals (fully occupied). $S_v$ does the same but excites at least one electron from the valence orbital \cite{Dutta2016}.  However, a class of triple excitations are included here in the estimations of the correlation energy through a perturbative approach. This formalism of the coupled-cluster method is well-known in the literature as RCCSD(T) \cite{Dutta2016}. By solving the energy eigen-value equations, we determine the amplitudes of excitations corresponding to the cluster operators and ionization energies for various valence configurations.
   
A normalized transition matrix element of any operator $\hat{O}$ can be expressed in a generalized form based on the RCC theory as
\clearpage
     \begin{eqnarray}
  O_{fi}&=& \frac{\langle\psi_f|\hat{O}|\psi_i\rangle}{\sqrt{\langle\psi_f|\psi_f\rangle \langle\psi_i|\psi_i\rangle}}\nonumber\\
  &=& \frac{\langle\phi_f| \{1+S^{\dagger}_f\}e^{T^\dagger}\hat{O}e^T \{1+S_i\}|\phi_i\rangle}{\sqrt{\langle\phi_f| \{1+S^{\dagger}_f\}e^{T^\dagger}e^T \{1+S_f\}|\phi_f\rangle\langle\phi_i|\{1+S^{\dagger}_i\}e^{T^\dagger}e^T \{1+S_i\}|\phi_i\rangle}}\nonumber\\
&=&
 \frac{1}{N}\big[\langle\phi_f|\bar{O} + (\bar{O}S_{1i} + S_{1f}^{\dagger}\bar{O}) + (\bar{O}S_{2i} + S_{2f}^{\dagger}\bar{O}) + ...|\phi_i\rangle\big].
     \end{eqnarray}
     Eq.(3) can be explicitly written in terms  of dressing of the open-shell operator on the closed-shell matrix element $\bar{O}$ (=$ e^{T^\dagger}\hat{O}e^T$), which in order to explicitly analyze the contributions from different kinds of correlation factors, appears as a consequence of the RCC theory\cite{Anal2017}. Here `1' and `2' in the subscript of the cluster operator indicate single and double excitation operators. The  difference between the corresponding matrix elements of $\bar{O}$ and $\hat{O}$ gives the core correlation contribution. Whereas, the matrix elements   $(\bar{O}S_{1i} + S_{1f}^{\dagger}\bar{O})$ and $(\bar{O}S_{2i} + S_{2f}^{\dagger}\bar{O})$ yield the pair correlation and core polarization, respectively, at the lowest order level.

     The detailed descriptions of the calculation of the electric dipole (\textit{E1}), magnetic dipole (\textit{M1}) and electric quadrupole (\textit{E}2) matrix elements and their associated transition oscillator strengths and probabilities   are available in Ref.  \cite{Dutta2016}. The lifetime $\tau_k$ of a state \textit{k} can be calculated using the formula, 
     \begin{equation}
     \tau_k = \left(\sum_i A_{k\rightarrow i} \right)^{-1},
     \end{equation}
    where $\sum_i A_{k\rightarrow i}$ sums the probabilities for all different channels of emissions from the state $k$  to all possible states $i$.

\section{RESULTS AND DISCUSSIONS}

We have used the basis set expansion technique to solve the Dirac-Fock equations for many-electron reference states in the RCC formalism, as mentioned in our earlier work \cite{Dixit2008, Dutta2014, Bhowmik2017, Dutta2011, Dutta2012, Dutta2013, Roy2014}. The reference states can be presented in the form of Slater determinants, i.e., antisymmetric products of the occupied DF orbitals. The Gaussian-type-orbital (GTO) bases are used \cite{Helgaker1995} to generate the radial part of the Dirac-Fock orbitals. Here we consider universal-type optimized exponent  in the radial part of the GTO basis functions. 
The numbers of the GTO basis functions for the $s, p, d, f, g$ and $h$-symmetries to obtain the DF orbitals are taken as $ 33, 30, 28, 25, 21$ and $ 20 $, respectively. The active orbitals for the RCC  calculations are chosen on the basis of convergence of core correlation energies.

Table 1 shows the calculated ionization energies (IEs) of different low-lying states for the ions considered here. Corresponding percentage (\%) deviations ('Dev') from the recommended reference data of the National Institute of Standards and Technology (NIST)\cite{NIST} are highlighted.  The average deviations are ~0.2\% for Y III, ~0.3\% for Zr IV, ~0.4\% for Nb V and  ~0.4\% for Mo VI. The IE values of six times ionized technetium, which is also naturally produced in  geological materials \cite{Dixon1997}, are estimated with a correlation-exhaustive many-body theory. There are many energy levels with more than one open shell present below the energy level of $5^2F$. Therefore, precise estimations of the $5^2F$ state and beyond are not possible within the present many-body method. Table 2 compares the fine structure splitting (FSS) values of the ground state ($4^2D$)  for the ions with results obtained from other calculations and experiments. The transitions within fine-structure states fall in the infrared regions and are important for astronomy, medical and physiological diagnostics. Our results show a  good agreement with the experimental  and empirical linear least squares fitted \cite{Ali1992} values. Table 2 also shows a definite trend where the RCC calculations produce FSS values larger than experimental ones and smaller than those of the multiconfiguration Dirac-Fock (MCDF) calculations\cite{Ali1992}.

\begin{figure}[!htbp]
\begin{center}$
\begin{array}{cc}
\includegraphics[width=90mm]{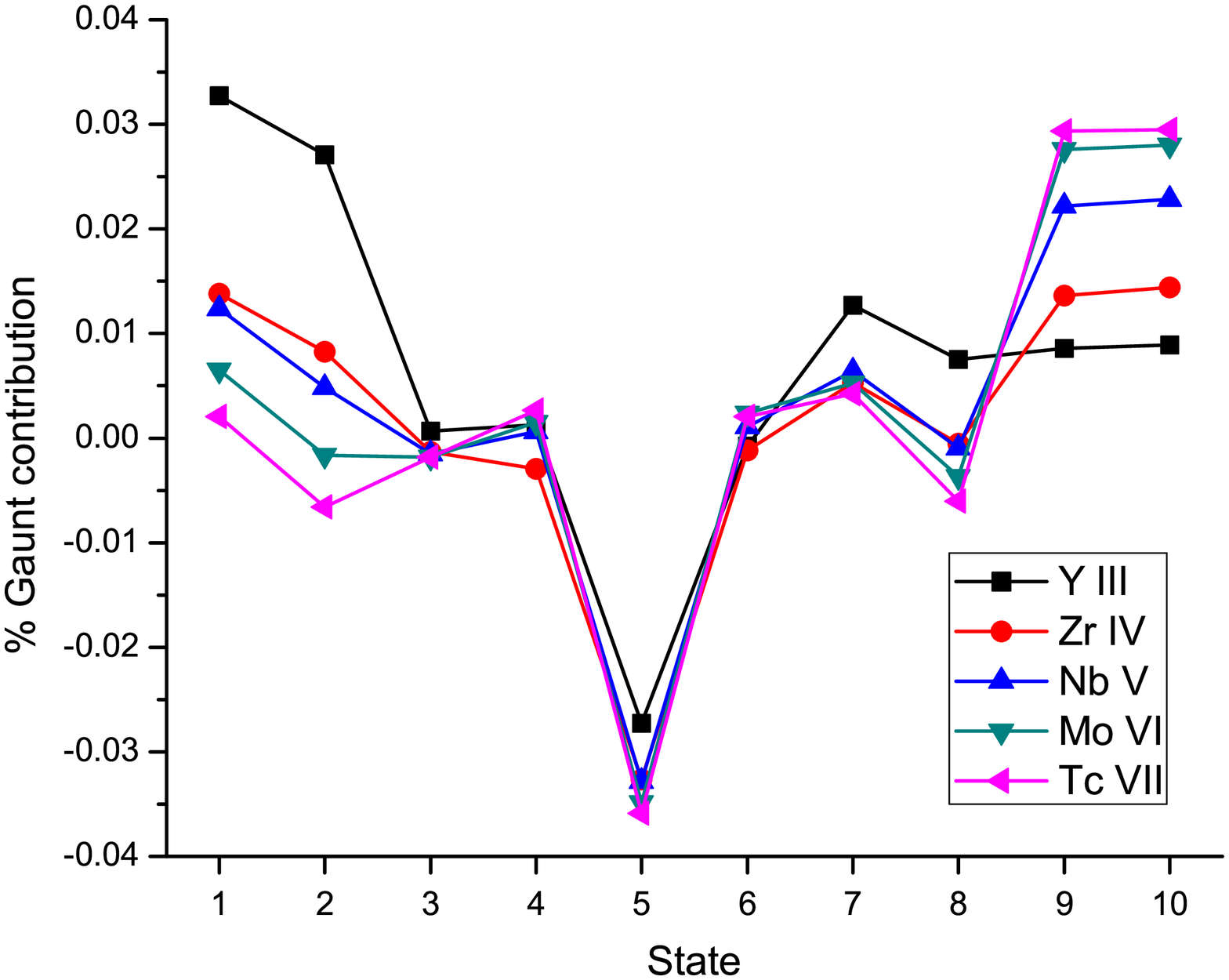}&
\includegraphics[width=90mm]{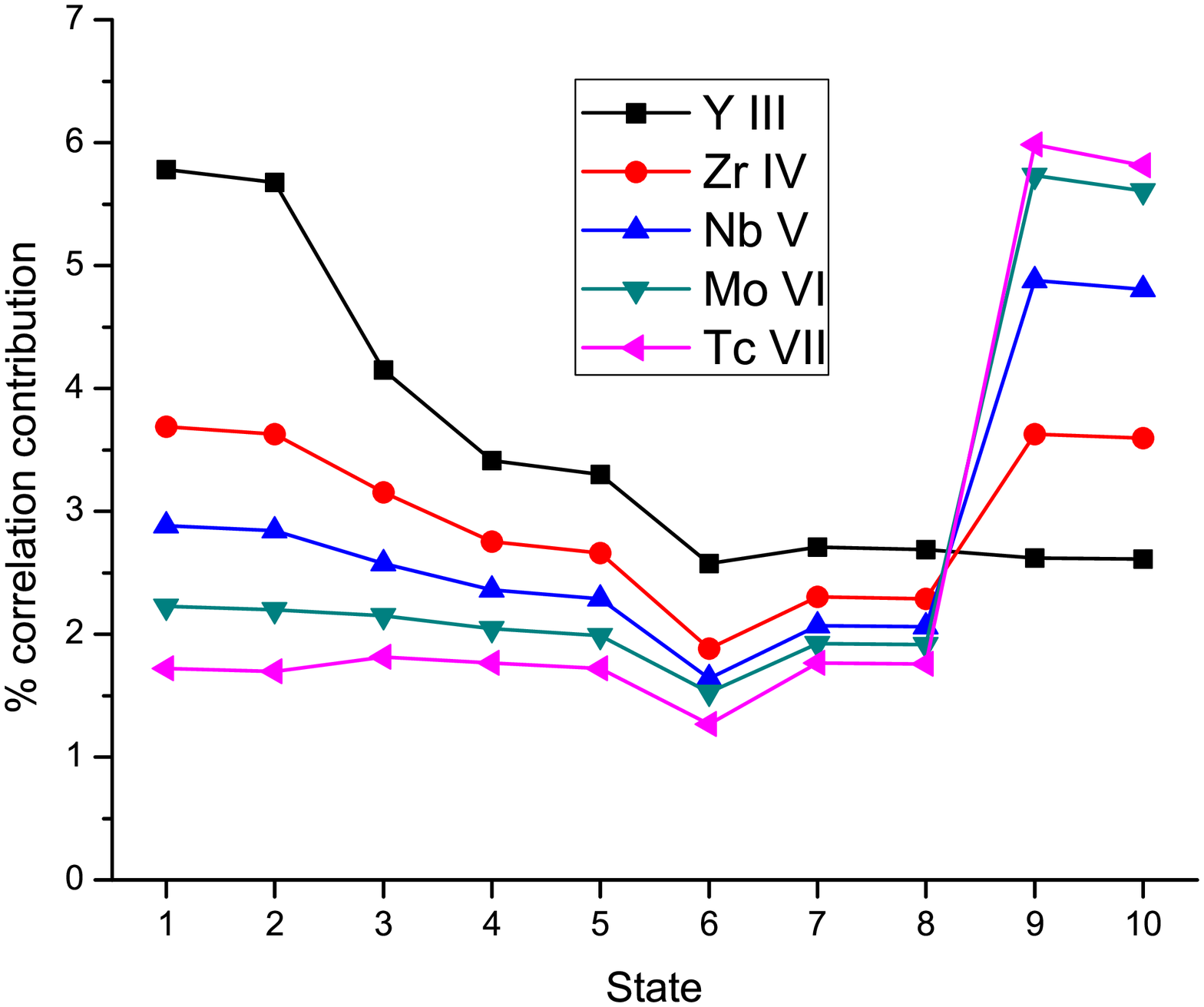}
\end{array}$
 (a) \hspace{8cm} (b)
\end{center}
\caption{Percentage Gaunt (a) and  correlation (b) contributions to the IEs . The numbers on the horizontal axis refer to the serial numbers of the different energy states given in the column "Sr. No" of Table 1.}

\label{A}
\end{figure}

\begin{figure}
\begin{center}$
\includegraphics[width=90mm]{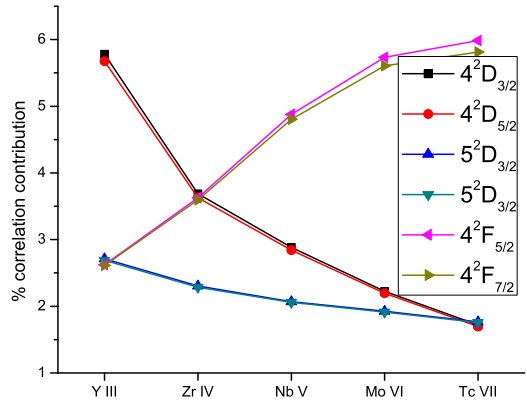}$
\end{center}
{\caption{Anomalous electronic correlation contribution to the energies of the $4^2F$ states compared to $4^2D$ and $5^2D$ states.}}
\label{B}
\end{figure}

         Figure~\ref{A}(a) and (b) show the percentage contributions of the Gaunt interaction \cite{Dutta2012,Breit1929} i.e., unretarded Breit interaction and Coulomb-correlation contribution to the IEs of states considered. For all the ions, the percentage Gaunt contributions vary within +0.04\% to -0.04\%.  It is interesting to note that for $4^2D$ states, Gaunt contributions are decreasing with the increase of ionization.
  The electron correlation contributes to the IEs very significantly (between 1\% and 6\%). In general, due to growing binding of the core and valence electrons to the nucleus with increasing ionization, the relative correlation contributions to the IEs for a particular state should decrease monotonically from Y III to Tc VII. Higher ionization means tighter valence electrons and hence relatively smaller impact of the Coulomb perturbation on the valence electrons. 
From Fig 1 (b), it is seen that this trend is obeyed well for all the states, except $4^2F$ states. It can be seen that the core-excited $4p^54d^2$ states have IEs close to the $4^2F$ states in the NIST data \cite{NIST}. This correlation contribution is estimated from the MCDF method using the GRASP92 \cite{Parpia2006} code just for investigation, and is found to be as small as 0.0013\%  of the orbital IE. The $4^2F$ states show a very different kind of response to electronic correlation with increasing ionization. The strong correlations in the $^2D$ and $^2F$ states change the ordering of the energy levels with changing degree of ionization. 
The IEs of the $4^2F$ states are increasing faster with the level of ionization compared to other states presented in the Table 1.
As a consequence, $4^2F$ states are above $5^2D$ for Y III, but appear just above $5^2P$ for Mo VI. Such a large correlation is primarily due to the amount of overlap between $5^2D$ and $4^2F$ wave functions. This  anomalous correlation contribution to $4^2F$ states with increasing ionization is shown in FIG~\ref{B}.

              In TABLE 3, we present the calculated Babushkin ($D_l$) and Coulomb ($D_v$) gauge values of matrix elements (amplitudes) for $E1$ transitions. Comparison of $D_l$ and $D_v$ values shows an overall good agreement, and this agreement is one of the measures of the accuracy of  calculations based on any many-electron method. It is known that the Coulomb gauge values are less stable compared to the corresponding Babushkin gauge values \cite{Grant2007}. Therefore, the Babushkin gauge values are commonly used for the calculations of astrophysically important parameters, such as oscillator strengths, transition rates and lifetimes. A large disagreement between the corresponding gauge values for $5^2D  \rightarrow 4^2F$ transition of Mo VI is observed. However, other transitions, whose initial or final state is either $5^2D$ or $4^2F$ state, show good agreement between the results obtained from the two gauges. Interestingly, even at the DF level, large  disagreement  is present for $5^2D  \rightarrow 4^2F$ transitions of Mo VI. The numerical DF calculations using  GRASP 92 Code \cite{Parpia2006} evaluate $f_{{Coulomb}}=0.001154$ and $f_{{Babushkin}}=0.17556 $. In Table 3, we have compared the $E1$ transition  matrix elements for Mo VI  with the available semiempirical results obtainted from NIST \cite{NIST, Reader2011}. The semiempirical results agree with the Babushkin values of matrix elements of  $E1$ transitions.

\begin{figure}

\begin{center}$
\begin{array}{cc}
\includegraphics[width=80mm]{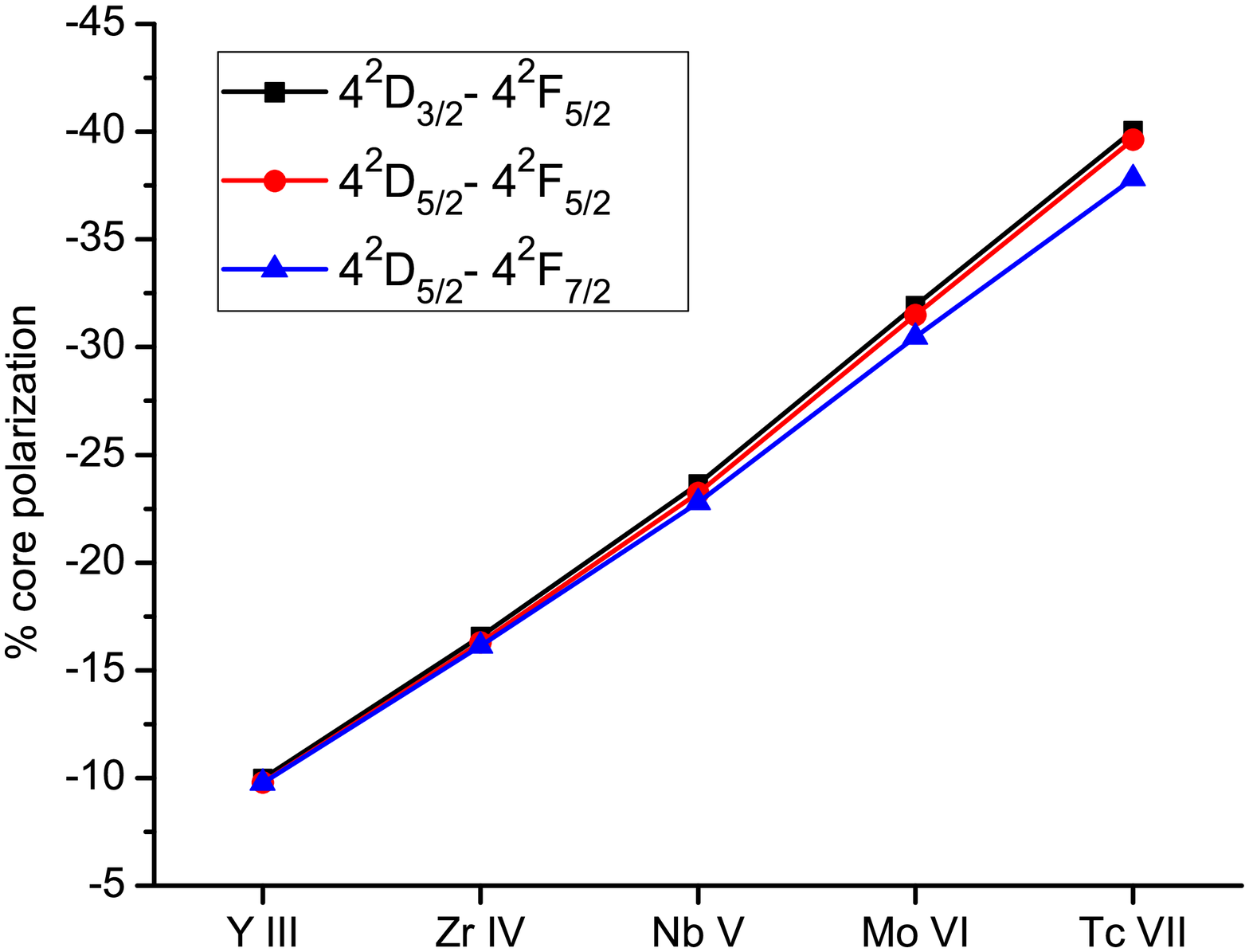}&
\includegraphics[width=80mm]{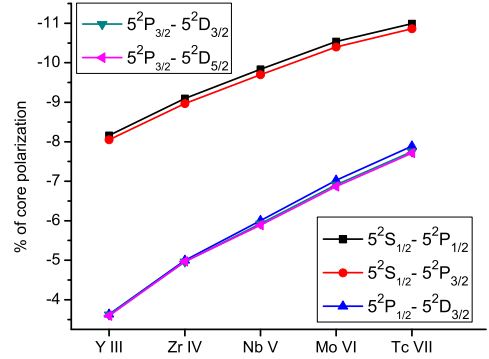}
\end{array}$
(a) \hspace{8cm} (b)
\end{center}
\begin{center}$
\begin{array}{cc}
\includegraphics[width=80mm]{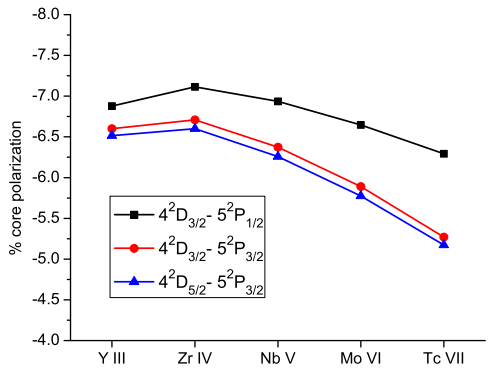}&
\includegraphics[width=80mm]{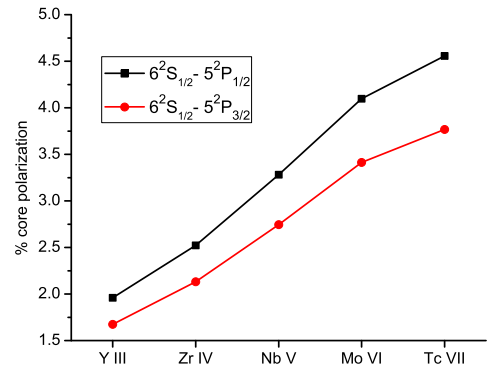}
\end{array}$
(c) \hspace{8cm} (d)
\end{center}
\begin{center}$
\includegraphics[width=80mm]{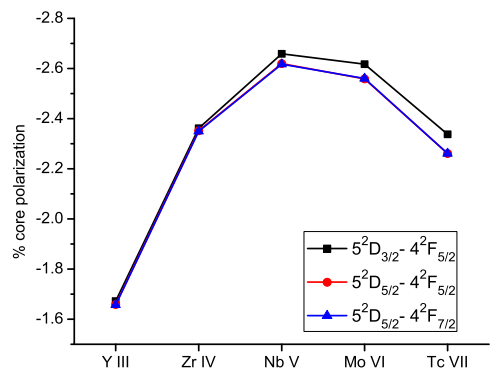}$
\end{center}
\begin{center}
(e)
\end{center}
\caption{Percentage core polarization  contributions with respect to DF  values for \textit{E1} transition matrix elements.}
\label{C}
\end{figure}

The calculated matrix elements of $E1$ transitions  have significant correlation contributions in the form of  core correlation (CC), core polarization (CP) and pair correlation (PC). There are a few higher-order non-negligible correlations, discussed in our earlier work \cite{Dutta2016}. They are considered here to compute the  matrix elements for $E1$ transitions. Since the  state-dependent effect of CP on $E1$ matrix elements is the dominant correlation mechanism \cite{Dutta2016, Anal2017}, we only present this effect graphically in various levels of ionization for all the transitions in Fig.~\ref{C}. In this figure, the percentage values of CP contributions to the Babushkin-gauge transition matrix elements are shown with respect to the corresponding values at the DF level. This correlation  effect is very strong for $4^2D-4^2F$  transitions and increases very rapidly with increasing ionization. The remarkably high CP contribution to the total correlation is observed for $4^2D-4^2F$ transitions of Mo VI (~30\%) and Tc VII (~40\%). With increasing ionization, the $5^2P-5^2D$ transitions follow the same pattern as  the $5^2S-5^2P$ transitions, but with smaller amplitudes.  For the $4^2D-5^2P$ transitions, Zr IV has the maximum CP contribution among the presented ions. The trend of the CP effect with the increasing atomic number (Z) for the  $6^2S-5^2P$ transitions is approximately the same as the trend for the $5^2S-5^2P$ transitions. However, the contributions are opposite in sign. The $5^2D-4^2F$ transitions are the most interesting transitions for the presented isoelectronic sequence due to the  collapse of the 5d and 4f orbitals. We observe the CP contribution of this transition to increase up to Nb V, and to decrease after this (i.e., for Mo VI and Tc VII). The $5^2D$ states are below $4^2F$ states for Y III (see Table 1), while they become closer in the next stage of ionization and are almost degenerate in Nb V. There is a cross-over between these two states at Mo VI. As a consequence, the $4^2F$ state lies below $5^2D$ for Mo VI and Tc VII. This cross-over  has a strong effect on the contribution of the CP to transition rates.

Table 4 shows the  oscillator strengths of $E1$ transitions along with the experimental and calculated RCC transition wavelengths. We have  compared our  results with the data available in literature. The oscillator strengths are calculated using the transition matrix elements in the Babushkin gauge obtained from Table 3 and the corresponding experimental and RCC wavelengths. Migdalek \cite{Migdalek2016} used two types of calculations, in which correlation effects were approximated with core polarization. The major difference between his two approaches is the different treatment of electron-electron exchange interaction. Similar non-local exchange terms have been included in our present calculations. Although the pair correlation (PC) terms are not as strong as CP, they are significant for precise calculations and are included in our present RCC formalism. Bi$\acute{e}$mont {\it et al.}\cite{Biemont2011} used core polarization correction as a modification of the relativistic Hartree-Fock method (HFR+CPOL) to calculate the oscillator strengths of various transitions of Y III. The all-order relativistic many-body perturbation theory (RMBPT) used by Safronova and Safronova \cite{Safronova2013}, employs correlation operators associated with single and double excitations, but with a linearized approximation. Therefore, their method is very similar to our RCC approach apart from  non-linear extra terms present in the RCC method. Their quoted uncertainty is around 0.5\% on average, whereas our estimated average uncertainty (as discussed further below) is 2.3\% including 2\% for other relativistic and correlation terms. Zhang et al. \cite{Zhang2009} has estimated a few oscillator strengths of Y III using a weakly bound electron potential method (WBEPM) where the parameters in the potential are fitted with experimental or other theoretical data.

           In TABLE 5, we present the Babushkin-gauge matrix elements of $E2$  transitions at the levels of the DF as well as the RCC method. The amplitudes of $E2$ transitions decrease with increasing ionization for each transition. The correlation effect reduces the values of RCC matrix elements compared to the DF values for most of the transitions. On average, the correlation contributions to the $E2$ amplitudes are around 10\%.
           Among them, the largest contributions are seen for transions between fine-structure states. Our RCC results for Y III are found to agree  with the  corresponding RMBPT results of Safronova and Safronova \cite{Safronova2013} within 0.3\% to 0.8\% and with the CCSD(T) results of Sahoo \textit{et al.} \cite{Sahoo2008} within 0.5\% to 0.6\%. Sahoo \textit{et al.} adopted the same version of the coupled-cluster theory that we have employed. Here the difference between the two coupled-cluster results may be due to choices of active orbitals and GTO basis.  
In the results of Sahoo et al., the calculated  pair correlation contributions with respect to DF  are around 5\% to 6\%, whereas in our case they are 7\% to 8\%. Therefore, proper choices of active orbitals and basis are important.

\begin{figure}

\begin{center}$
\begin{array}{cc}
\includegraphics[width=60mm]{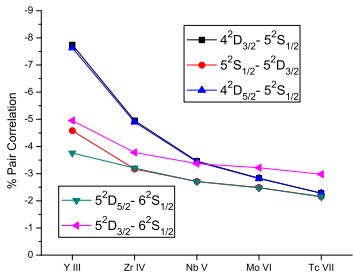}
\includegraphics[width=60mm]{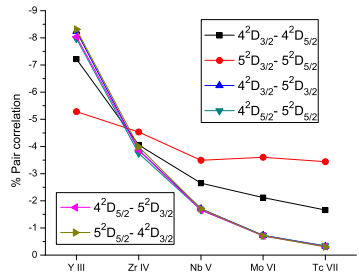}&
\includegraphics[width=60mm]{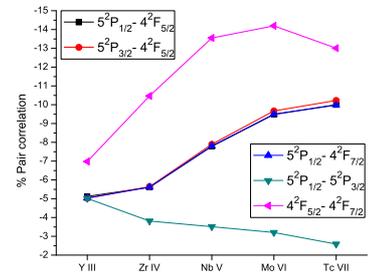}
\end{array}$
  (a) \hspace{7cm}  (b) \hspace{6cm} (c)
\end{center}

\caption{Percentage pair correlation  contributions with respect to DF  value for \textit{E2} transition matrix elements.}
\label{D}
\end{figure}

           As the PC contribution dominates over the CP contribution for the $E2$ transitions, we graphically present only the percentage of the PC contribution with respect to DF in Fig.~\ref{D}. The PC contribution decreases smoothly with the increase of ionic charges for all the transitions except the transitions associated with $4^2F$ states. The contribution varies from 1\% to 8\% of the DF values in extreme cases. However, for the transitions associated with  $4^2F$ states, it varies from 6\% to 14\%.

 The $M1$ transition amplitudes of the ions are tabulated in Table 6 and compared with the only other available calculations\cite{Safronova2013,Sahoo2008} for the $4^2D_\frac{3}{2}\rightarrow4^2D_\frac{5}{2}$ and $4^2D_\frac{3}{2}\rightarrow5^2S_\frac{1}{2}$ transitions of Y III.  As in the case of the $E2$ transition, our RCC results are in better agreement with the RMBPT estimations by Safronova and Safronova \cite{Safronova2013} compared to the CCSDpT results of Sahoo \textit{et al.} \cite{Sahoo2008}.  As usual, the dominant transitions here are those between the fine-structure states of the same term. For them, $M1$ transition amplitudes are larger than $E2$ transition amplitudes. Some of these transitions can be crucial for density estimation in various stellar and interstellar media.

The accuracy of our calculations of these transition amplitudes can be further estimated by comparing our calculated lifetimes, as seen in Table 7, with the results available in the literature. Maniak \textit{et al.} \cite{Maniak1994} measured the lifetimes of $5^2P_\frac{1}{2}$ and $5^2P_\frac{3}{2}$ of Y III using beam foil spectroscopy. Our calculated lifetimes agree (within 9\% for $5^2P_\frac{1}{2}$ and 5\% for $5^2P_\frac{3}{2}$) with the experimental lifetimes from \cite{Maniak1994}. Those authors have also calculated these lifetimes using the method of Coulomb approximation with a Hartree-Slater model core. Another set of data for these transitions are measured by the time-resolved laser induced fluorescence method \cite{Biemont2011}. We find a better agreement (1\% for $5^2P_\frac{1}{2}$ and 5\% for $5^2P_\frac{3}{2}$) between our calculated lifetimes and the experimental lifetimes from \cite{Biemont2011}. The estimated lifetimes of  Safronova and Safronova\cite{Safronova2013} are in good agreement with our results with the average discrepancy of 0.8\%.

Theoretical  uncertainties in the calculated parameters depend on the quality of the generated wave functions, as  the amplitudes are significant at the DF levels. The uncertainties are calculated from the root-mean-square deviation of transition amplitudes, which are calculated with the help of  orbital wave functions obtained from the GTO basis and   a sophisticated numerical approach (GRASP92 code) \cite{Parpia2006}. We should also consider the other correlation terms and quantum electro-dynamic  effect  (totally at most $\pm$2\%),  which are not considered in this paper.  Considering all these, the maximum estimated uncertainties for allowed transition amplitudes are $\pm$2.8\%, $\pm$2.5\%, $\pm$2.9\%, $\pm$2.7\%, $\pm$2.5\% for Y III, Zr IV, Nb V, Mo VI, Tc VII respectively. Whereas, maximum estimated uncertainties for forbidden transition amplitudes are $\pm$3.4\%, $\pm$3.6\%, $\pm$4.3\%, $\pm$5.3\%, $\pm$3.5\% for the above ions respectively. We do not consider the $5^2D-4^2F$ transitions for Mo VI in the accuracy calculations due to the above-mentioned large disagreement between the Babushkin and Coulomb gauge values of oscillator strengths at the Dirac-Fock level for these transitions.

\clearpage

\section{Conclusion}
 
We have performed accurate calculations of the ionization energies of different low-lying states of a few Rb-like ions and several transition parameters between these states using RCC method. We report the effect of the energy state crossing within the iso-electronic sequence.  Correlation study of  these parameters shows interesting phenomena of core polarization and pair correlations. Calculated transition parameters are compared with existing theoretical and experimental results, wherever available, and good agreement is found in most cases. Many of the transition properties of Tc VII are evaluated for the first time in this literature. The accuracy of the calculations is estimated and compared with other published calculations. The calculated transition matrix elements and lifetimes are important in estimation of  abundance of atomic and ionic elements in astrophysical bodies  through line detection in Earth- and space-based telescopes.
\section*{ACKNOWLEDGMENTS}
The calculations were carried out in the IBM cluster at IIT-Kharagpur, India, funded by DST-FIST (SR/FST/PSII-022/2010).

\clearpage
 \begin{table}[!htbp]
 \scriptsize
  \centering
 \caption{RCC calculated ionization energies (in cm$^{-1}$) of ground and low-lying excited states  are presented along with the  percent of the deviations  ("Dev") from corresponding experimental values (NIST\cite{NIST}). } 
 \begin{tabular}{l c c c c c c c c c c }
 \hline\hline
Sr. & State & \multicolumn{2}{c}{Y III} & \multicolumn{2}{c}{Zr IV} & \multicolumn{2}{c}{Nb V} & \multicolumn{2}{c}{Mo VI} &{Tc VII}
\\
 No. &  & RCC& Dev& RCC& Dev& RCC& Dev& RCC& Dev& RCC
 \\
 \hline
 1& $4^2D_{\frac{3}{2}}$ & 165030.02	&	-0.3	&	276481.42	&	-0.4	&	407569.40	&	-0.1	&	555177.10	&	0.0	&	718455.74 \\
 2 & $  4^2D_{\frac{5}{2}}$ &164267.73	&	-0.3	&	275177.17	&	-0.4	&	405626.79	&	-0.1	&	552498.32	&	0.0	&	714936.68\\
 3 & $  5^2S_{\frac{1}{2}}$ &158311.28	&	0.2	&	239422.92	&	0.0	&	331931.95	&	0.0	&	435684.30	&	0.1	&	550080.50
\\
4 &   $5^2P_{\frac{1}{2}}$ & 124112.55	&	0.0	&	195410.90	&	-0.1	&	278216.76	&	-0.2	&	372395.20	&	-0.1	&	477209.30
\\
 5 & $  5^2P_{\frac{3}{2}}$ &122537.61	&	0.0	&	192910.76	&	-0.1	&	274552.65	&	-0.2	&	367362.62	&	-0.1	&	470667.03
\\
 6 & $   6^2S_{\frac{1}{2}}$ & 78725.28	&	-0.1	&	124715.94	&	-0.3	&	177383.98	&	-1.1&	239314.30	&	-0.8	&	310434.72
\\
 7 & $   5^2D_{\frac{3}{2}}$ &76866.88	&	-0.4	&	130623.91	&	-0.2	&	195687.39	&	-0.3	&	271250.46	&	-0.4	&	357919.22
\\
 8 &  $  5^2D_{\frac{5}{2}}$ &76662.71	&	-0.4	&	130251.95	&	-0.3	&	195136.98	&	-0.3	&	270440.78	&	-0.4	&	356798.20
\\
 9 & $   4^2F_{\frac{5}{2}}$ & 64280.45	&	-0.3	&	117804.35	&	-0.6	&	191032.82	&	-0.8	&	285041.73	&	-1.1	&	398493.03
\\
10 &   $  4^2F_{\frac{7}{2}}$ & 64286.54	&	-0.3	&	117803.33	&	-0.6	&	190969.75	&	-0.8	&	284796.54	&	-1.0	&	397950.46
\\ 

  \hline

\end{tabular}
\label{I}
\end{table}  

\begin{table}
\scriptsize
  \centering
\caption{Comparison of ground state fine-structure splittings (FSS) (in cm$^{-1})$  with other theoretical and experimental results. Fitted$\rightarrow $ Linear least squares fitted values\cite{Ali1992}, Exp$\rightarrow $ Experimental data from NIST\cite{NIST}}     
 \begin{tabular}{c c c c c c c}
 \hline
  System & \multicolumn{2}{c}{Our} &  \multicolumn{2}{c}{Ali\cite{Ali1992}}&  Exp\cite{NIST} & Fitted\cite{Ali1992}\\
 \hline
    &  DF  &  RCC  & DF  & MCDF &  &  \\
    \hline
 YIII& 570 & 762 &&&724.15 &\\
 ZrIV& 1111 &1304&1125&1182&1250.7&1249\\
 NbV & 1753 & 1943&1746&1798&1867.4 &1865\\
 MoVI& 2500 & 2679&2469&2517&2583.5 &2584\\
 TcVII&3355 & 3519&3292&3347&&3414\\
 \hline
 
\end{tabular}
\label{II}
\end{table}

\begin{table}[!htbp]
 \scriptsize
 \caption{Absolute values of Babushkin  ($D_l$) and Coulomb ($D_v$) gauge matrix elements for different \textit{E1} transitions in a.u.  } 
 \begin{tabular}{c c c c c c c c c c c c}
 \hline\hline
Transition & \multicolumn{2}{c}{Y III} & \multicolumn{2}{c}{Zr IV} & \multicolumn{2}{c}{Nb V} & \multicolumn{3}{c}{Mo VI} & \multicolumn{2}{c}{Tc VII}\\
&$D_l$&$D_v $&$D_l$&$D_v $&$D_l$&$D_v $&$D_l$&$D_v $&NIST&$D_l$&$D_v $ \\
\hline
$4^2D_{\frac{3}{2}} \rightarrow 5^2P_{\frac{1}{2}}$ &1.9566	&	1.4340	&	1.4615	&	1.1832	&	1.1825	&	1.0024	&	0.9947	&	0.8102&1.0163 &	0.8583	&	0.7407
	\\
               \hspace{.7cm}$\rightarrow 5^2P_{\frac{3}{2}}$ &0.8622	&	0.6357	&	0.6406	&	0.5190	&	0.5160	&	0.4349	&	0.4329	&	0.3654& 0.4544 &	0.3728	&	0.3260
	\\
              \hspace{.7cm}$\rightarrow 4^2F_{\frac{5}{2}}$ &2.2530	&	2.2541	&	2.0317	&	2.0332	&	1.7996	&	1.8125	&	1.4943	&	1.6284	& 1.2604&	1.1983	&	1.3178
\\
\hline
 $4^2D_{\frac{5}{2}} \rightarrow 5^2P_{\frac{3}{2}}$ &2.6210	& 1.9200 &	1.9508	& 1.5720 &	1.5738	&	1.3200	&	1.3219	&	1.1210&	1.3677&	1.1399	&	1.1009	\\
              \hspace{.7cm}$\rightarrow  4^2F_{\frac{5}{2}}$ &0.6104	&	0.6091	&	0.5502	&	0.5489	&	0.4874	&	0.4904	&	0.4046	&	0.4371	& 0.3379&	0.3243	&	0.3549	\\
               \hspace{.7cm} $\rightarrow  4^2F_{\frac{7}{2}}$ & 2.7314&2.7270& 2.4656&2.4598&2.1922
&2.2035&1.8380&1.9735&1.5784&1.4967&1.6183\\
 \hline
 $5^2S_{\frac{1}{2}}\rightarrow5^2P_{\frac{1}{2}}$&2.5552	&	2.4944	&	2.2184	&	2.1460	&	1.9708	&	1.8960&1.7708	&	1.7614
& 1.8621	&	1.6200	&	1.6080
	\\
                \hspace{.7cm} $\rightarrow5^2P_{\frac{3}{2}}$&3.6140	&	3.5166	&	3.1396	&	3.0256	&	2.7914	&	2.6722
	&	2.5091	&	2.4705
& 2.6270&	2.2951	&	2.2550	\\
 \hline
 $5^2P_{\frac{1}{2}}\rightarrow5^2D_{\frac{3}{2}}$ &3.8649	&3.7247	&3.4312	&3.3346	&3.0750	&2.9676	&2.7571	&2.7247& 2.9946
	&2.5104	&2.4880	
	\\
                 \hspace{.7cm} $\rightarrow6^2S_{\frac{1}{2}}$&	1.7241	&1.6443	&1.3405	&1.2801	&1.1193	&1.0676	&0.9926	&0.92895&0.9683 &0.8791	&0.8472
	\\
\hline
$5^2P_{\frac{3}{2}}\rightarrow5^2D_{\frac{3}{2}}$&1.7752	&1.7103&	1.5718	&1.5284	&1.4088	&1.3619&1.3404	&1.2629	&1.2351	&1.1481	&1.1324	\\
                 \hspace{.7cm} $\rightarrow5^2D_{\frac{5}{2}}$&5.3050	&5.1052	&4.6981	&4.5646	&4.2167	&4.0727	&3.7815	&3.6895&4.0185	&3.4388	&3.3761		\\
                  \hspace{.7cm} $\rightarrow6^2S_{\frac{1}{2}}$& 2.5584	&2.4385	&1.9997	&1.9108	&1.6791	&1.6042	&1.4955 &	1.4129&1.3688&1.3287	&1.2937
 \\
 \hline
 $5^2D_{\frac{3}{2}} \rightarrow4^2F_{\frac{5}{2}}$ & 8.1236&	8.3437&	5.6716	&6.1963	&4.0627	&4.9701	&2.9421	&0.4703&2.9052 &2.2188	&1.6401
	\\
 \hline
 $5^2D_{\frac{5}{2}} \rightarrow4^2F_{\frac{5}{2}}$  &2.2353	&2.2353	&1.5130	&1.6593	&1.0838	& 1.3422	&0.7837&	0.0342&0.7763
	&0.5902	&0.4416	\\
                  \hspace{.7cm} $\rightarrow4^2F_{\frac{7}{2}}$ &9.7167	&9.9939	&6.7678	&7.4230	&4.8536	&6.0213	&3.5166	&0.2182&	3.4400&2.6543	&1.9901	\\
\hline

\end{tabular}

\label{III}
\end{table}


\begin{table}
\scriptsize
\caption{Comparison of oscillator strengths or f-values (in a.u.) with other results. $\lambda_{Exp}$ \& $\lambda_{RCC}$ represent the NIST \& RCC values of vacuum wavelength (in nm), respectively.
$f_{RCC}$ are calculated with both wavelengths.}

\centering 
\begin{tabular}{c c c  c c c l c}
\hline\hline
Ion &Transition & $\lambda_{Exp}$\cite{NIST} & $\lambda_{RCC}$&\multicolumn{2}{c}{$f_{RCC}$} &$f_{others}$&$f_{NIST}$\\
&&&&$\lambda_{Exp}$&$\lambda_{RCC}$&&\\
\hline
&5S-5P&&&&&&\\
Y III &1/2-1/2& 294.7 & 292.4&  0.337& 0.339	& 0.333$^a$, 0.332$^b$	&\\
 &&  &   	&&	&	 0.353$^c$, 0.360$^d$, 0.399$^e$	&\\

Zr IV&1/2-1/2& 228.7 & 227.2& 0.327	&0.329 &	0.331$^a$, 0.393$^e$&\\
Nb V&1/2-1/2& 187.7 & 186.2& 0.314	&0.317 &	0.326$^a$, 0.385$^e$&
\\
Mo VI&1/2-1/2& 159.5 & 158.0 &0.299	& 0.301& 0.299$^f$, 0.376$^e$ &0.330
\\
Tc VIII&1/2-1/2& & 137.2 &  &0.290	& 0.367$^e$	&\\
Y III  &1/2-3/2&  281.9 &279.5 & 0.704 &0.710 & 0.696$^a$, 0.695$^b$ &\\
  &  & 	&	&&	&	 0.738$^c$, 0.751$^d$, 0.831$^e$ &\\
Zr IV &1/2-3/2&  216.4 & 215.0 & 0.692	& 0.696 &	0.699$^a$, 0.829$^e$ &
	\\
Nb V&1/2-3/2 & 175.8 & 174.3& 0.673	& 0.679&	0.697$^a$, 0.821$^e$ &
	\\
Mo VI &	1/2-3/2 & 147.9& 146.4 &0.646 & 0.653 &0.646$^f$, 0.810$^e$	&0.709\\
Tc VII	&1/2-3/2 &  &125.9 & & 0.635 & 0.798$^e$ &\\

&4D-5P&&&&\\
Y III&3/2-1/2&241.5& 244.4 &0.120& 0.119 &	0.120$^a$,	0.119$^b$& \\
& &	&	&&	& 0.127$^c$, 0.138$^e$ & \\
Zr IV&3/2-1/2&122.0 & 123.3 &0.132 &0.131 &	0.132$^a$, 0.157$^e$ &\\
Nb V&3/2-1/2&77.4 & 77.3& 0.137& 0.137 &0.133$^a$, 0.161$^e$ &\\
Mo VI&3/2-1/2 &54.8	&54.7 &	0.137& 0.137	&0.133$^f$, 0.160$^e$ &0.143\\
Tc VII&3/2-1/2&	& 41.5&	&0.135	& 0.157$^e$ &\\
Y III&3/2-3/2&232.8 & 235.3 & 0.024	& 0.024&	0.024$^a$,0.024$^b$&\\
&& 	&	&&	& 0.026$^c$, 0.0278$^e$ &\\
Zr IV&3/2-3/2&118.4& 119.7& 0.026 &	0.026&	0.026$^a$, 0.0308$^e$	&\\
Nb V&3/2-3/2& 75.3& 75.2 &0.027	& 0.027 &0.026$^a$, 0.0311$^e$&\\
Mo VI	&3/2-3/2&53.4 &	53.2 &0.027 & 0.027& 0.026$^f$, 0.0305$^e$&0.029
\\
Tc VII	&3/2-3/2&	&40.4 &&0.026	& 0.0296$^e$ &\\
Y III&5/2-3/2&236.8	& 239.6&0.147&	0.145&	0.145$^a$	,0.145$^b$&\\
&	&	&	&&	&0.155$^c$ &\\
Zr IV&5/2-3/2& 120.2& 121.6&0.160 &	0.158 &0.157$^a$	&\\
Nb V&5/2-3/2&76.4& 76.3& 0.164	& 0.164& 0.157$^a$&\\
Mo VI	& 5/2-3/2&	 54.1& 54.0& 0.163&	0.164&	0.161$^f$&0.175\\

Tc VII	&5/2-3/2&	&40.9 && 0.161	&	&\\ 
&5D-4F&&&&&&\\
Y III&3/2-5/2& 786.7& 794.5&0.637 & 0.631&	0.655$^a$, 0.615$^b$&\\
&	&	&	&&	&	 0.674$^c$, 0.635$^d$&\\
Zr IV&3/2-5/2& 805.5& 780.1&0.303	& 0.313&	0.442$^a$&\\
Nb V&3/2-5/2&2805 & 2148.4 & 0.045	& 0.058&	0.176$^a$&\\
Mo VI &3/2-5/2&	633.8 & 725.1&	0.069& 0.060 &	&0.0674\\
Tc VII	&3/2-5/2& & 246.5 &	&0.101	&	&\\
Y III&5/2-5/2& 799.2 & 807.6& 0.030	& 0.031&	0.031$^a$, 0.029$^b$&\\
&&	&	&&	&	 0.032$^c$, 0.030$^d$&\\
Zr IV&5/2-5/2& 828.9 & 803.4& 0.014	&0.014	&0.021$^a$&\\
Nb V&5/2-5/2&3310.5 & 2436.6 &0.002	& 0.002&	0.008$^a$&\\
Mo VI	&5/2-5/2&603.7	& 684.9&0.005 &	0.005 &		&0.005\\
Tc VII	&5/2-5/2&	& 239.8 && 0.003	&	&\\
Y III&5/2-7/2& 799.4 & 808.0& 0.598	& 0.592&	0.616$^a$, 0.579$^b$ &\\
&& 	&	&&	&	 0.633$^c$	0.661$^d$ &\\
Zr IV&5/2-7/2& 827.5 & 803.3 &0.280	&0.289 &	0.413$^a$&\\
Nb V&5/2-7/2& 3165.5& 2399.7& 0.038	& 0.050&	0.163$^a$&\\
Mo VI&5/2-7/2&	619& 696.6& 0.076 &	0.067&	&0.0724\\
Tc VII	&5/2-7/2&	& 243.0 &	& 0.110	&	&\\
&4D-4F&&&&\\
Y III&3/2-5/2& 98.9& 99.3& 0.390 &	0.388& 0.398$^a$,0.388$^b$&\\
&& 	&	&&	& 0.385$^c$, 0.478$^e$&\\
Zr IV&3/2-5/2&62.9& 63.0 &0.499	& 0.497 &0.502$^a$, 0.676$^e$&\\
Nb V&3/2-5/2&46.5& 46.2 &0.529 & 0.533&	0.613$^a$, 0.877$^e$&\\
Mo VI	&3/2-5/2&37.4& 37.0 & 0.453	& 0.458& 0.291$^f$, 0.290$^g$, 1.023$^e$ &0.322	\\
Tc VII & 3/2-5/2& &31.3 & &0.349	& 1.101$^e$ &	\\
Y III&5/2-5/2&99.6& 100.0 &0.019 &	0.019&	0.019$^a$, 0.019$^b$, 0.018$^c$ &\\
Zr IV&5/2-5/2&63.4& 63.5 &0.024	& 0.024& 0.024$^a$&\\
Nb V&5/2-5/2& 46.9& 46.6 & 0.026 &	0.026& 0.030$^a$&\\
Mo VI&5/2-5/2 & 37.8 &37.4 & 0.022	&0.022 & 0.015$^f$,	 0.014$^g$  &0.015\\ 
Tc VII	&5/2-5/2 & &31.6 &	&0.017	&		&\\

Y III&5/2-7/2&99.6 & 100.0 &0.379 &	0.378 &0.385$^a$, 0.378$^b$,  0.364$^c$ &\\
Zr IV&5/2-7/2& 63.4 &63.5 & 0.486 &	0.484& 0.485$^a$&\\
Nb V&5/2-7/2& 46.8 & 46.6 &0.520 & 0.522&	0.590$^a$&\\
Mo VI	&5/2-7/2&37.5 & 37.4 &0.453	& 0.458 &&0.333\\
Tc VII	&5/2-7/2& &31.5
 &&0.360	&	&\\
\hline

\end{tabular}\\
\begin{tiny}
\hspace{-0.1cm}a$\Rightarrow$(RMP+EX+CP) Ref. \cite{Migdalek2016}, b$\Rightarrow$(RMBPT) Ref.\cite{Safronova2013}, c$\Rightarrow$(HFR+CPOL) Ref.\cite{Biemont2011}.
\end{tiny}\\
\begin{tiny}
\hspace{-0.4cm} d$\Rightarrow$ (WBEPM) Ref. \cite{Zhang2009}, e$\Rightarrow$ (DF) Ref.\cite{Zilitis2007}, f$\Rightarrow$ (MCDHF) Ref. \cite{Fischer2011}, g$\Rightarrow$ (RCI) Ref. \cite{Pan2006}
\end{tiny}\\

\label{IV}
\end{table}

\begin{table}
\scriptsize
  \centering
\caption{ Absolute values of matrix elements  for different $E2$ transitions along with the Dirac Fock values (DF). All the results are in a.u.}  
 \begin{tabular}{c c c c  c c c c c c c}
 \hline\hline
Transition & \multicolumn{2}{c}{Y III} & \multicolumn{2}{c}{Zr IV} & \multicolumn{2}{c}{Nb V} & \multicolumn{2}{c}{Mo VI} & \multicolumn{2}{c}{Tc VII}\\
& DF &RCC &DF &RCC &DF &RCC &DF &RCC  &DF&RCC \\
\hline
$4^2D_{\frac{3}{2}}\rightarrow4^2D_{\frac{5}{2}}$&3.5399	&	3.1403	& 	2.3822	&	2.1458	&	1.8026	&	1.6313	&	1.4427	&	1.3013	&	1.1970	&	1.0775	\\
& &3.114(9)\cite{Safronova2013} 
&&&&&&&&\\
&& 3.1555\cite{Sahoo2008} &&&&&&&&\\

               \hspace{.75cm}$\rightarrow5^2S_{\frac{1}{2}}$&6.6962	&	6.1006	&	4.3587	&	4.0825	&	3.1495	&	2.9955	&	2.4161	&	2.3128	&	1.9286	&	1.8573	\\
& &6.08(2)\cite{Safronova2013} 
&&&&&&&&\\
&&6.1364\cite{Sahoo2008}
&&&&&&&&\\
                   
                 \hspace{.75cm}$\rightarrow6^2S_{\frac{1}{2}}$&1.6872	&	1.4559	&	0.5068	&	0.4968	&	0.1338	&	0.1663	&	0.0033	&	0.0412	&	0.0470	&	0.0120	\\
              \hspace{.75cm}$\rightarrow5^2D_{\frac{3}{2}}$&4.1954	&	3.8700	&	2.6540	&	2.5947	&	1.8840	&	1.8987	&	1.4605	&	1.4978	&	1.2059	&	1.2494	\\

                \hspace{.75cm}$\rightarrow5^2D_{\frac{5}{2}}$&	2.72177&2.53489	&	1.7199	&	1.6789	&	1.2196	&	1.2283	&	0.9443	&	0.9677	&	0.7794	&	0.8068	\\

$4^2D_{\frac{5}{2}}\rightarrow5^2S_{\frac{1}{2}}$&8.2786	&	7.5575	&	5.3988	&	5.0644	&	3.9061	&	3.7202	&	2.9997	&	2.8754	&	2.3968	&	2.3115	\\

&& 7.52(2)\cite{Safronova2013} 
&&&&&&&&\\
&& 7.6003\cite{Sahoo2008}
&&&&&&&&\\
                 \hspace{.75cm} $\rightarrow6^2S_{\frac{1}{2}}$&2.1269	&	1.8473	&	0.6531	&	0.6405	&	0.1835	&	0.2225	&	0.0179	&	0.0629	&	0.0472	&	0.0058	\\
                 \hspace{.75cm} $\rightarrow5^2D_{\frac{3}{2}}$&2.7950	&	2.5860	&	1.7717	&	1.7343	&	1.2595	&	1.2702	&	0.9771	&	1.0026	&	0.8073	&	0.8369	\\
              \hspace{.75cm}$\rightarrow5^2D_{\frac{5}{2}}$	&	5.5403	&5.1230&		3.5082	&3.4314	&		2.4914	&2.5127	&		1.9304	&1.9810		&	1.5946&	1.6531
\\
  $5^2D_{\frac{5}{2}}\rightarrow5^2D_{\frac{3}{2}}$ &22.2480	&	20.9643	&	13.6725	&	12.9653	&	9.7018	&	9.2812	&	7.4497	&	7.1040	&	5.7577	&	5.4901	\\

$5^2S_{\frac{1}{2}}\rightarrow5^2D_{\frac{3}{2}}$ &11.2285	&	10.7102	&	9.0124	&	8.6449	&	7.4555	&	7.1748	&	6.2349	&	6.0045	&	5.2636	&	5.0785	\\

               \hspace{.75cm}$\rightarrow5^2D_{\frac{5}{2}}$&
               13.6890	&	13.0547	&	10.9984	&	10.5457	&	9.1047	&	8.7624	&	7.6183	&	7.3377	&	6.4344	&	6.2088	\\

$6^2S_{\frac{1}{2}}\rightarrow5^2D_{\frac{3}{2}}$&35.7613	&	33.8451	&	22.2873	&	21.3356	&	15.4609	&	14.8510	&	11.4344	&	10.9901	&	8.7166	&	8.3948	\\

                   \hspace{.75cm} $\rightarrow5^2D_{\frac{5}{2}}$&44.0116	&	41.6730	&	27.4718	&	26.2338	&	19.0753	&	18.3284	&	14.1224	&	13.5784	&	10.7763	&	10.3820	\\
  $5^2P_{\frac{1}{2}}\rightarrow5^2P_{\frac{3}{2}}$&15.9451	&	14.9795	&	11.4961	&	10.9055	&	8.9628	&	8.5111	&	7.1318	&	6.7820	&	5.8477	&	5.5886	\\
 $5^2P_{\frac{1}{2}}\rightarrow4^2F_{\frac{7}{2}}$& 30.2896	&	28.5567	&	20.7062	&	19.2940	&	14.6874	&	13.2451	&	10.4048	&	9.0773	&	7.5616	&	6.4392	\\
 
$4^2F_{\frac{5}{2}}\rightarrow5^2P_{\frac{1}{2}}$ &22.5766	&	21.2601	&	15.5071	&	14.4446	&	11.0391	&	9.9509	&	7.8493	&	6.8400	&	5.7293	&	4.8680	\\
                    \hspace{.75cm} $\rightarrow5^2P_{\frac{3}{2}}$&12.3659	&	11.6582	&	8.4552	&	7.8756	&	5.9986	&	5.4013	&	4.2493	&	3.6950	&	3.0874	&	2.6168	\\
                   \hspace{.75cm} $\rightarrow4^2F_{\frac{7}{2}}$& 18.1675	&	16.8276	&	9.3217	&	8.2510	&	5.4510	&	4.5946	&	3.4028	&	2.7752	&	2.3167	&	1.8522	\\

   \hline
\end{tabular}

\label{V}
\end{table}

\begin{table}[!htbp]
  \centering
\caption{ Calculated matrix elements (absolute values) of  \textit{M1} transitions are presented in a.u. The numbers in brackets represent the power of 10.}  
\resizebox{\textwidth}{!}{\begin{tabular}{c c c c c c c c c c c}
 \hline\hline
Transition & \multicolumn{2}{c}{Y III} & \multicolumn{2}{c}{Zr IV} & \multicolumn{2}{c}{Nb V} & \multicolumn{2}{c}{Mo VI }& \multicolumn{2}{c}{Tc VII}\\
& DF & RCC & DF & RCC & DF & RCC & DF & RCC & DF & RCC\\
\hline
 
$4^2D_{\frac{3}{2}}\rightarrow4^2D_{\frac{5}{2}}$  & 1.5490	&	1.5493	&	1.5489	&	1.5492	&	1.5489	&	1.5491	&	1.5488	&	1.5490	&	1.5488	&	1.5489	
\\
 
 && 1.5491$^a$ &&&&&&&&   
 \\
 &&  1.5434$^b$ &&&&&&&&   
 \\
                   \hspace{0.75cm} $\rightarrow5^2S_{\frac{1}{2}}$& 2.557[-06]	&	2.307[-05]	&	4.202[-06]	&	1.574[-05]	&	1.133[-05]	&	1.871[-05]	&	2.710[-05]	&	1.829[-06]	&	4.804[-05]	&	1.849[-05]	
 \\
 
 & & 3.950[-05]$^a$ &&&&&&&&   
 \\
 & &  3.337[-04]$^b$ &&&&&&&&   
 \\
               \hspace{0.75cm} $\rightarrow6^2S_{\frac{1}{2}}$& 1.744[-05]	&	2.628[-05]	&	1.208[-05]	&	2.405[-05]	&	1.622[-05]	&	4.011[-05]	&	9.375[-06]	&	3.488[-05]	&	1.229[-05]	&	1.167[-05]	
 \\
 
                \hspace{0.75cm}$\rightarrow5^2D_{\frac{3}{2}}$& 2.029[-04] & 2.493[-03]& 3.054[-04] &7.117[-03] & 4.11[-04] & 1.080[-02]	&  5.354[-04] &   7.360[-03]	& 6.677[-04] &	5.675[-03] 
\\
              \hspace{0.75cm} $\rightarrow5^2D_{\frac{5}{2}}$&6.629[-03]   &8.508[-3]  & 3.054[-04] &  7.093[-03] &  7.692[-03]& 2.847[-03]  &  8.182[-03] & 5.119[-03] 	&	8.679[-03]  &	6.517[-03]	
\\
$4^2D_{\frac{5}{2}}\rightarrow5^2D_{\frac{3}{2}}$&6.822[-03] &6.509[-03]&   7.583[-03]	& 1.173[-02] &    8.045[-03] &  1.399[-02]	&	8.636[-03]  &  	1.284[-02]	  &  9.239[-03] & 1.260[-02] 
\\
 
   $5^2D_{\frac{5}{2}}\rightarrow5^2D_{\frac{3}{2}}$& 1.5491	&	1.5492	&	1.5491	&	1.5491	&	1.5491	&	1.5491	&	1.5490	&	1.5491	&	1.5490	&	1.5491	
\\
 
  $5^2P_{\frac{1}{2}}\rightarrow5^2P_{\frac{3}{2}}$&1.1542	&	1.1541	&	1.1541	&	1.1540	&	1.1539	&	1.1539	&	1.1538	&	1.1538	&	1.1537	&	1.1537
\\
$5^2P_{\frac{3}{2}}\rightarrow4^2F_{\frac{5}{2}}$ & 2.494[-05]	&	2.739[-05]	&	3.108[-05]	&	9.080[-06]	&	3.437[-05]	&	3.229[-05]	&	2.728[-05]	&	2.232[-05]	&	2.408[-05]	&	3.948[-05]	
\\

$4^2F_{\frac{5}{2}}\rightarrow4^2F_{\frac{7}{2}}$ & 1.8516	&	1.8517	&	1.8516	&	1.8517	&	1.8516	&	1.8516	&	1.8515	&	1.8510	&	1.8515	&	1.8495	
\\
 
\hline

\end{tabular}}
\begin{tiny}
a$\rightarrow$ (RMBPT) Ref.\cite{Safronova2013},
\end{tiny}
\begin{tiny}
b$\rightarrow$ (CCSDpT) Ref.\cite{Sahoo2008}
\end{tiny}
\label{VI}
\end{table}

\begin{table}[!htbp]

\caption{ Our RCC calculated life times of few low-lying states are compared with the values available in the literature.}

\resizebox{\textwidth}{!}{\begin{tabular}{c c c c c c c c c}
 \hline
Ions & \multicolumn{2}{c}{$4^2D_{\frac{5}{2}}$(Sec)} & \multicolumn{2}{c}{$5^2S_{\frac{1}{2}}$(Sec)}  & \multicolumn{2}{c}{$5^2P_{\frac{1}{2}}$(nSec)  } & \multicolumn{2}{c}{$5^2P_{\frac{3}{2}}$(nSec)}  \\
&RCC&Other&RCC&Other&RCC&Other&RCC&Other\\
\hline
Y$^{2+}$& 244.08 & 244.1$^a$,  & 10.76 & 10.85$^a$,  & 1.874& 1.898$^a$, 1.94$^c$  &1.702&  1.723$^a$, 1.76$^c$ 	\\
&  & 245.89$^b$ &  & 10.63$^b$ & & 2.146$^d$  &  & 1.948$^d$ \\
&  &  &  &  & &  1.9(10)$^e$ && 	1.8(20)$^e$\\
&  &  &  &  & &   2.06(8)$^f$ && 1.79(8)$^f$	\\

Zr$^{3+}$& 47.38& &5.67$\times 10^{-03}$& & 0.622&0.588$^c$&0.5786 &0.550$^c$\\
 Nb$^{4+}$& 12.65&&3.34$\times 10^{-04}$&&0.274&0.249$^c$&0.259&0.238$^c$\\
 Mo$^{5+}$& 5.38&& 5.69$\times 10^{-05}$&&0.146&0.130$^c$&0.139&0.126$^c$
\\
Tc$^{6+}$& 2.13&&1.60$\times 10^{-05}$&&0.0869&0.0768$^c$&0.083&0.0753$^c$
\\
\hline

\end{tabular}}
\label{VII}
\begin{tiny}
  $a$ $\Rightarrow$ (RMBPT) Ref. \cite{Safronova2013},
  $b$ $\Rightarrow$ (CCSDpT) Ref. \cite{Sahoo2008},
$c$ $\Rightarrow$ (DF) Ref. \cite{Zilitis2007},
\end{tiny}

\begin{tiny}
$d$ $\Rightarrow$ (Coulomb approximation with Hartree-Slater Core) Ref. \cite{Maniak1994},
$e$ $\Rightarrow$ (Time-revolved laser-induced fluorescence)Ref. \cite{Biemont2011},
\end{tiny}

\begin{tiny}
$f$ $\Rightarrow$ (Beam foil spectroscopy) Ref. \cite{Maniak1994}
\end{tiny}
\end{table}

\clearpage




\end{document}